\documentclass[11pt, paper=a4]{scrartcl}

\usepackage{listings}
\usepackage[dvipsnames]{xcolor}
\usepackage{pifont}
\usepackage{amsmath}
\usepackage{bm}
\usepackage[margin=2.5cm,includehead=true,includefoot=true,centering,]{geometry}
\usepackage{footnotebackref}
\usepackage[sort&compress, numbers]{natbib}
\usepackage[semibold, scale=0.9]{sourcecodepro}
\usepackage{dirtree}
\usepackage{multicol}
\usepackage{graphicx}
\usepackage{authblk}

\setkomafont{disposition}{\normalfont\bfseries}
\DeclareOldFontCommand{\bf}{\normalfont\bfseries}{\mathbf}

\setcapindent{0em}

\usepackage{hyperref}
\hypersetup{                   %
    colorlinks = true,         
    linkcolor  = MidnightBlue, 
    urlcolor   = BrickRed,     
    citecolor  = MidnightBlue, 
    pdfcreator={LaTeX via pandoc with the Eisvogel template}
}

\makeatletter
\AtBeginDocument{%
  \newcounter{llabel}[lstlisting]%
  \renewcommand*{\thellabel}{%
    \ifnum\value{llabel}<0 %
      \@ctrerr
    \else
      \ifnum\value{llabel}>10 %
        \@ctrerr
      \else
        \protect\ding{\the\numexpr\value{llabel}+201\relax}%
      \fi
    \fi
  }%
}

\newlength{\llabelsep}
\newcommand*{\llabel}[1]{%
  \begingroup
    \refstepcounter{llabel}%
    \label{#1}%
    \llap{%
      \thellabel\kern\llabelsep
      \hphantom{\lst@numberstyle\the\lst@lineno}%
      \kern\lst@numbersep
    }%
  \endgroup
}
\makeatother

\usepackage{caption}
\DeclareCaptionFont{white}{\color{white}}
\DeclareCaptionFont{white}{\color{white}}
\DeclareCaptionFormat{listing}{\colorbox{gray}{\parbox{\dimexpr\linewidth-2\fboxsep\relax}{#1#2#3}}}
\definecolor{listing-background}{HTML}{F4F4F4}
\definecolor{listing-background-terminal}{HTML}{000000}
\definecolor{listing-rule}{HTML}{B3B2B3}
\definecolor{listing-numbers}{HTML}{B3B2B3}
\definecolor{listing-text-color}{HTML}{000000}
\definecolor{listing-text-color-terminal}{HTML}{FFFFFF}
\definecolor{listing-keyword}{HTML}{1284CA}
\definecolor{listing-keyword-2}{HTML}{1800AA} 
\definecolor{listing-keyword-3}{HTML}{9137CB} 
\definecolor{listing-identifier}{HTML}{435489}
\definecolor{listing-string}{HTML}{00999A}
\definecolor{listing-comment}{HTML}{8E8E8E}

\definecolor{draculabg}      {RGB} {40,   42,   54}
\definecolor{draculacl}      {RGB} {68,   71,   90}
\definecolor{draculafg}      {RGB} {248,  248,  242}
\definecolor{draculacomment} {RGB} {98,   114,  164}
\definecolor{draculacyan}    {RGB} {139,  233,  253}
\definecolor{draculagreen}   {RGB} {24,   182,  82}
\definecolor{draculaorange}  {RGB} {255,  184,  108}
\definecolor{draculapink}    {RGB} {255,  121,  198}
\definecolor{draculapurple}  {RGB} {189,  147,  249}
\definecolor{draculared}     {RGB} {255,  85,   85}
\definecolor{draculayellow}  {RGB} {241,  250,  140}

\lstdefinelanguage{mypython}{
language=Python,
keywordstyle     = [1]{\color{draculapurple}\bfseries},
keywordstyle     = [2]{\color{draculapink}},
keywordstyle     = [3]{\color{draculacyan}},
keywordstyle     = [6]{\color{draculared}\bfseries\itshape},
identifierstyle  = \color{draculacl},
commentstyle     = \color{draculacomment},
stringstyle      = \color{draculaorange},
showstringspaces = false,
morekeywords     = [1]{as, assert, nonlocal, with, yield, self, True, False, None}, 
morekeywords=[2]{__init__,__add__,__mul__,__div__,__sub__,__call__,__getitem__,__setitem__,__eq__,__ne__,__nonzero__,__rmul__,__radd__,__repr__,__str__,__get__,__truediv__,__pow__,__name__,__future__,__all__}, 
morekeywords=[2]{object,type,isinstance,copy,deepcopy,zip,enumerate,reversed,list,set,len,dict,tuple,range,xrange,append,execfile,real,imag,reduce,str,repr}, 
morekeywords=[2]{ode,fsolve,sqrt,exp,sin,cos,arctan,arctan2,arccos,pi, array,norm,solve,dot,arange,isscalar,max,sum,flatten,shape,reshape,find,any,all,abs,plot,linspace,legend,quad,polyval,polyfit,hstack,concatenate,vstack,column_stack,empty,zeros,ones,rand,vander,grid,pcolor,eig,eigs,eigvals,svd,qr,tan,det,logspace,roll,min,mean,cumsum,cumprod,diff,vectorize,lstsq,cla,eye,xlabel,ylabel,squeeze}, 
morekeywords=[2]{import_chains,prior,plot_posteriors,S,spectrum, spec_importer, log_spectrum, g_rho, g_s, compute_bf, plot_chains},
morekeywords=[6]{Exception,NameError,IndexError,SyntaxError,TypeError,ValueError,OverflowError,ZeroDivisionError}, 
}
\lstdefinelanguage{mybash}{
language=bash,
keywordstyle     = {\color{draculapurple}\bfseries},
keywordstyle     = [1]{\color{draculapink}\bfseries},
keywordstyle     = [2]{\color{draculapurple}},
keywordstyle     = [3]{\color{draculaorange}},
morekeywords     = [1]{pull, run, create, install,notebook, shell},
morekeywords     = [2]{docker, singularity, conda, bash, pip, brew, curl, sudo, apt,get,jupyter},
}
\lstdefinestyle{eisvogel_listing_style}{
  numbers          = left,
  xleftmargin      = 2.25em,
  xrightmargin      = \fboxsep,
  framexleftmargin = 2em,
  backgroundcolor  = \color{listing-background},
  basicstyle       = \color{draculacl}\linespread{1.0}%
                      \ttfamily{},
  numberstyle       = \color{draculacl}\small%
                      \ttfamily{},
  escapeinside     = {/*@}{@*/}, 
  breaklines       = true,
  sensitive        = true,
  frame            = single,
  framesep         = 0.19em,
  rulecolor        = \color{listing-rule},
  frameround       = ffff,
  tabsize          = 4,
  numbers          = none,
  aboveskip        = 1.0em,
  belowskip        = 0.5em,
  abovecaptionskip = 0em,
  belowcaptionskip = 0.75em,
  literate         =
  {á}{{\'a}}1 {é}{{\'e}}1 {í}{{\'i}}1 {ó}{{\'o}}1 {ú}{{\'u}}1
  {Á}{{\'A}}1 {É}{{\'E}}1 {Í}{{\'I}}1 {Ó}{{\'O}}1 {Ú}{{\'U}}1
  {à}{{\`a}}1 {è}{{\`e}}1 {ì}{{\`i}}1 {ò}{{\`o}}1 {ù}{{\`u}}1
  {À}{{\`A}}1 {È}{{\`E}}1 {Ì}{{\`I}}1 {Ò}{{\`O}}1 {Ù}{{\`U}}1
  {ä}{{\"a}}1 {ë}{{\"e}}1 {ï}{{\"i}}1 {ö}{{\"o}}1 {ü}{{\"u}}1
  {Ä}{{\"A}}1 {Ë}{{\"E}}1 {Ï}{{\"I}}1 {Ö}{{\"O}}1 {Ü}{{\"U}}1
  {â}{{\^a}}1 {ê}{{\^e}}1 {î}{{\^i}}1 {ô}{{\^o}}1 {û}{{\^u}}1
  {Â}{{\^A}}1 {Ê}{{\^E}}1 {Î}{{\^I}}1 {Ô}{{\^O}}1 {Û}{{\^U}}1
  {œ}{{\oe}}1 {Œ}{{\OE}}1 {æ}{{\ae}}1 {Æ}{{\AE}}1 {ß}{{\ss}}1
  {ç}{{\c c}}1 {Ç}{{\c C}}1 {ø}{{\o}}1 {å}{{\r a}}1 {Å}{{\r A}}1
  {€}{{\EUR}}1 {£}{{\pounds}}1 {«}{{\guillemotleft}}1
  {»}{{\guillemotright}}1 {ñ}{{\~n}}1 {Ñ}{{\~N}}1 {¿}{{?`}}1
  {…}{{\ldots}}1 {≥}{{>=}}1 {≤}{{<=}}1 {„}{{\glqq}}1 {“}{{\grqq}}1
  {”}{{''}}1
}
\newcommand{\ptarcade}{\texttt{PTArcade}}

\definecolor{mygreen}{RGB}{31, 138, 112}
\newcommand{\pref}[3]{\hyperlink{#1}{\hypertarget{#3}{{\color{mygreen}#2}}}}
\newcommand{\preff}[3]{\hyperlink{#1}{\hypertarget{#3}{{#2}}}}
\def\mat#1{\boldsymbol{#1}}
\newcommand\abhb{A_{\scriptscriptstyle\textrm{BHB}}}
\newcommand\gbhb{\gamma_{\scriptscriptstyle\textrm{BHB}}}

\begin{document}

\title{\texttt{\href{https://andrea-mitridate.github.io/PTArcade}{\color{black}PTArcade \\\vspace{-10pt}{\Large v0.1.0}}}}

\newcommand{\TR}[1]{\textcolor{red}{[Tobias+Richard: #1]}}
\newcommand{\AM}[1]{\textcolor{blue}{[Andrea: #1]}}
\newcommand{\DW}[1]{\textcolor{green}{[David: #1]}}
\hyphenation{bootstrap}
\hyphenation{ptarcade}

\date{}

\setlength{\affilsep}{-0.05em}
\author{
\textbf{ Main developers:}\vspace{-1em}\and 
{\normalsize Andrea Mitridate}\\ {\footnotesize \emph{Deutsches Elektronen-Synchrotron DESY, Notkestr. 85, D-22607 Hamburg, Germany}}\vspace{-1em}\and
{\normalsize David Wright}\\ {\footnotesize\emph{Department of Physics, University of Central Florida, Orlando, FL 32816-2385, USA}}\and
\textbf{Contributors:\vspace{-1em}}
\and{\normalsize  Richard von Eckardstein and Tobias Schr\"{o}der}\\ {\small\emph{Institute for Theoretical Physics, University of M\"{u}nster, 48149 M\"{u}nster, Germany}}\vspace{-1em}\and
{\normalsize Jonathan Nay}\\ {\footnotesize\emph{Department of Physics, The University of Texas at Austin, Austin, TX 78712, USA}}\vspace{-1em}\and 
{\normalsize Ken Olum} \\ {\footnotesize\emph{Institute of Cosmology, Department of Physics and Astronomy, Tufts University, Medford, MA 02155, USA}}\vspace{-1em}\and 
{\normalsize Kai Schmitz} \\ {\footnotesize\emph{Institute for Theoretical Physics, University of M\"{u}nster, 48149 M\"{u}nster, Germany}}\vspace{-1em}\and
{\normalsize Tanner Trickle} \\ {\footnotesize\emph{Theoretical Physics Division, Fermi National Accelerator Laboratory, Batavia, IL 60510, USA}}}

\maketitle
\begin{abstract}
    This is a lightweight manual for \ptarcade{}, a wrapper of \texttt{ENTERPRISE} and \texttt{ceffyl} that allows for easy implementation of new-physics searches in PTA data. In this manual, we describe how to get \ptarcade{} installed (either on your local machine or an HPC cluster). We discuss how to define a stochastic or deterministic signal and how \ptarcade{} implements these signals in PTA-analysis pipelines. Finally, we show how to handle and analyze the \ptarcade{} output using a series of utility functions that come together with \ptarcade{}.
\end{abstract}
\clearpage
\tableofcontents

\section{Introduction}
The detection of gravitational waves (GWs) by the LIGO and VIRGO collaborations \cite{LIGOScientific:2016aoc} heralds the beginning of GW astronomy. The extremely weak interaction between GWs and matter makes them ideal probes for dense astrophysical and cosmological environments, one relevant example being the pre-recombination Universe. This epoch in the cosmological evolution is characterized by high densities of charged particles, which makes it opaque to electromagnetic radiation. However, any gravitational-wave signal produced during this epoch would propagate essentially unimpeded over cosmic distances to be measured today. Detecting a primordial GW signal will then provide a direct glimpse into the primordial Universe and potentially allow us to test beyond Standard Model (BSM) physics, where the production of primordial GWs is a ubiquitous feature~\cite{Maggiore:1999vm,Caprini:2018mtu,Christensen:2018iqi}. 

Recently, several pulsar timing array (PTA) collaborations have found convincing evidence for a gravitational wave background (GWB) in the nanohertz band (see for example reference~\cite{aaa+23_gwb}). The origin of this background is still unknown, and while supermassive black hole binaries (SMBHBs) remain the primary suspect~\cite{Rahagopal+Romani-1995,Jaffe:2002rt,Wyithe+Loeb-2003,Sesana:2004sp,Burke-Spolaor:2018bvk,aaa+23_smbh}, a primordial origin is also a viable explanation at this stage. The recent NANOGrav search for new-physics signals \cite{aaa+23_newphys} has considered several BSM models that could generate a primordial GWB compatible with the one observed in the NANOGrav 15yr dataset. However, many models remain to be tested. 

In this work, we present \ptarcade{}, a wrapper of \texttt{ENTERPRISE} \cite{2019ascl.soft12015E, enterprise} and \texttt{ceffyl}~\cite{lamb2023need}.
PTArcade aims to provide an accessible way to perform Bayesian analyses of new-physics signals with PTA data. The user can either specify the signal by providing the GW energy density spectrum as a fraction of the closure density $\Omega_{\mathrm{GW}}(f;\,\vec\theta)$ (in case of GWB signals) or the signal time series $h(t, \vec\theta)$ (in case of deterministic signals). Here, $\vec\theta$ is an array containing all the model parameters characterizing the new-physics signal for which \ptarcade{} will allow to derive posterior probability distributions and upper limits.

Users can specify their models by using simple Python files, which allows for great flexibility and allows to specify signals either analytically or using tabulated data. \ptarcade{} is shipped with sensible default settings that closely resemble official PTA analyses, but users may override them on a case-by-case basis through a pure-Python configuration file.

While all the necessary information needed to install and run \ptarcade{} can be found in this manual, more details and examples can be found in the \ptarcade{} \href{https://andrea-mitridate.github.io/PTArcade/}{documentation web page}.

\section{Quick Start}\label{sec:quick}
The first step consists in installing \ptarcade{}. The easiest way to do this is to use the \texttt{Python} package manager \href{https://conda.io/projects/conda/en/latest/index.html}{\texttt{conda}}. Simply download \href{https://andrea-mitridate.github.io/PTArcade/assets/downloads/ptarcade.yml}{this environment file}, open a terminal, and type\footnote{Here we are assuming that the \texttt{.yml} file is located in the local directory. If that is not the case you should pass the full path to the \texttt{.yml} file when executing \texttt{conda env create}.}
\begin{lstlisting}[language=mybash]
conda env create -f ptarcade.yml
\end{lstlisting}
If everything went smoothly, great, \ptarcade{} is now installed on your machine and you can start using it!\footnote{If you encounter any problem during the installation, refer to the "Troubleshooting" section in the \ptarcade{} \href{https://andrea-mitridate.github.io/PTArcade}{documentation web page} for possible solutions.} 

To guide our discussion, we will consider a toy model: Suppose you have a model that produces a GWB with a broken power-law spectrum of the form
\begin{align}\label{eq:gwb_ex}
    h^2\Omega_{\scriptscriptstyle\mathrm{GW}}(f; A_*,f_*)=A_*\left(\frac{f_*}{f}+\frac{f}{f_*}\right)^{-1}\,,
\end{align}
and you want to know for what values of the parameters $A_*$ and $f_*$ this GWB can reproduce the signal observed in the NG15 data. The first step is to create, what we call, a \emph{model file}. This is a simple Python file that contains the definition of the GWB spectral shape and the prior distributions for the model parameters. For the GWB from our example, the model file is:
\begin{lstlisting}[language=mypython, 
title={A Typical Model File},
numbers=left,
mathescape,
style=eisvogel_listing_style,
label=Code:Ex_ModelFile]
from ptarcade.models_utils import prior

parameters = {'log_A_star' : prior("Uniform", -14, -6),
              'log_f_star' : prior("Uniform", -10, -6)} $\llabel{model.parameters}$

def S(x):
    return 1 / (1/x + x)

def spectrum(f, log_A_star, log_f_star): $\llabel{model.spectrum}$
    A_star = 10**log_A_star
    f_star = 10**log_f_star
    
    return A_star * S(f/f_star)
\end{lstlisting}
In general, the model file for a GWB signal needs to contain the following two things:
\begin{itemize}
    \item[\ref{model.parameters}]\texttt{parameters}: This variable has to be assigned to a dictionary whose keys are strings corresponding to the model parameters' names and whose values are the parameters' priors. The prior distributions can be specified using the \texttt{prior} function from the \texttt{models\_utils} module. The syntax for this function is the following: The first argument that is passed to this function is a string specifying the prior type. The subsequent arguments are the parameters describing the prior.\footnote{A list of the built-in prior types and their parameters can be found in Table~\ref{tab:priors_tab}.} By default, the parameters are assumed to be common across all pulsars in the array. If you want a parameter to be pulsar dependent, you need to add the flag \texttt{common=False} when instantiating the \texttt{prior} object.

    In our example, we have chosen the names \texttt{log\_A\_star} and \texttt{log\_f\_star} for the parameters  $\log_{10} A_*$ and $\log_{10}(f_*/{\textrm Hz})$.  For both, we have chosen uniform priors with the ranges $[-14,-6]$ and $[-10, -6]$, respectively.
    
    \item[\ref{model.spectrum}]\texttt{spectrum(f, ...)}: this function specifies the GWB spectrum. Its first parameter should be named \texttt{f}, and it has to be a NumPy array containing the frequencies (in units of Hz) at which the spectrum is evaluated. The names of the remaining parameters should match the ones defined in the \texttt{parameters} dictionary (in our case \texttt{log\_A\_star} and \texttt{log\_f\_star}). The \texttt{spectrum} function should return a NumPy array with the same dimensions as \texttt{f} and contain the value of $h^2\Omega_{\scriptscriptstyle\mathrm{GW}}$ at each of the frequencies in \texttt{f}. 
\end{itemize}

Once you have created the model file, you are ready to run \ptarcade{}. To proceed, open a terminal window and type the following:
\begin{lstlisting}[language=mybash,
numbers=left,
mathescape]
ptarcade -m ./model_file.py 
\end{lstlisting}
The argument passed to the \texttt{-m} input flag is supposed to be the path to the model file. By default, \ptarcade{} will output a chain of Markov Chain Monte Carlo samples (together with other files discussed in the next section) to the directory \texttt{./chains/np\_model/chain\_0}.\footnote{The user can change the location of the output directory. See Section~\ref{sec:details} for more details on how to do this.} This chain of MC samples can then be used to derive the posterior distribution for the parameters of the user-specified signal.

\begin{figure}[t]
    \centering
    \includegraphics[width=0.60\textwidth]{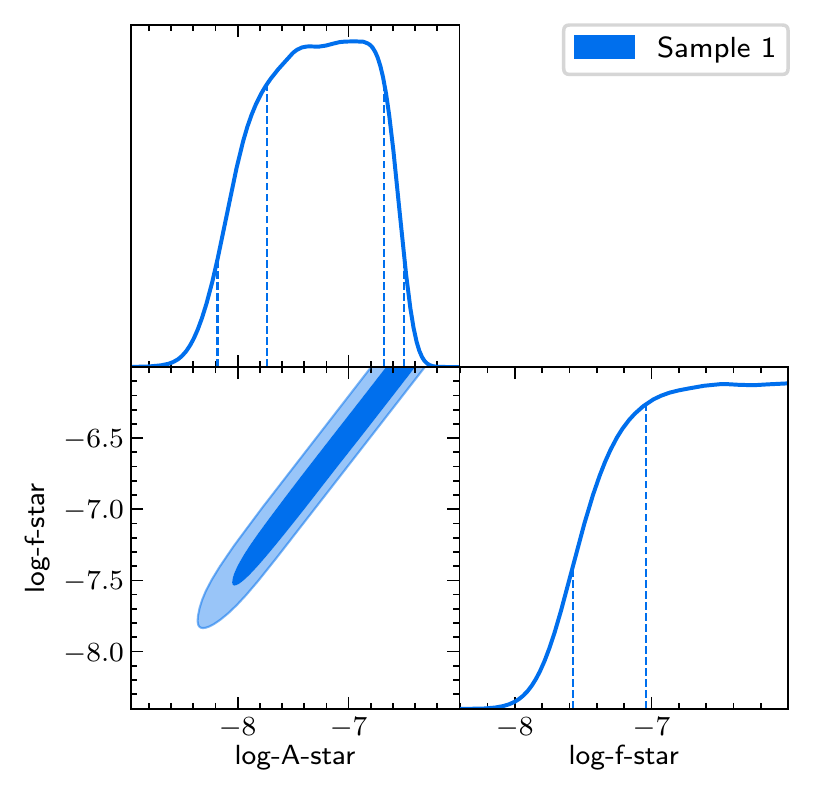}
    \caption{Reconstructed posterior distributions for the parameters of the example GWB described in Eq.~\eqref{eq:gwb_ex}. On the diagonal of the corner plot, we report the 1D marginalized distributions together with the $68\%$ and $95\%$ Bayesian credible intervals (vertical lines), while the off-diagonal panels show the $68\%$ (darker) and $95\%$ (lighter) Bayesian credible regions in the 2D posterior distributions. We construct all credible intervals and regions by integrating over the regions of highest posterior density.}
    \label{fig:posterior_ex}
\end{figure}

Once the sampling of the chains is complete, you can use one of the many market-available tools to produce posterior-distribution plots. Two popular choices are \href{https://corner.readthedocs.io/en/latest/}{\texttt{corner}}~\cite{corner} and \href{https://getdist.readthedocs.io/en/latest/intro.html}{\texttt{GetDist}}~\cite{Lewis:2019xzd}. \ptarcade{} itself provides two modules, \texttt{chains\_utils} and \texttt{plot\_utils}, that can be used to help in this procedure (see section \ref{subsec:utils}). When using these two modules, producing the posterior plots  can be done with only a few lines of code:
\begin{lstlisting}[
title={Loading Chains and Parameter Files},
language=mypython,
numbers=left,
mathescape,
style=eisvogel_listing_style,
label=Code:Ex_PostPlot,
]
import ptarcade.plot_utils as p_utils
import ptarcade.chains_utils as c_utils

chain, params = c_utils.import_chains('./chains/np_model/') $\llabel{chains_import}$

p_utils.plot_posteriors([chain] , [params]) $\llabel{chains_plotting}$
\end{lstlisting}
In \ref{chains_import}, we use the function \hyperlink{utils.import-chains}{\color{mygreen}\texttt{import\_chains}} to load the MCMC chains and the parameter files of the run. This function takess the path to a folder containing the chains and loads them into a NumPy array. Additionally, it returns a dictionary containing the names of the model parameters as keys and a list with the parameters' prior ranges as values. The \hyperlink{utils.plot-posteriors}{\color{mygreen}\texttt{plot\_posteriors}} function in \ref{chains_plotting} produces a plot with the 2D and 1D marginalized posteriors for the parameters of our model. For the example at hand, we obtain the plot which is shown in Figure~\ref{fig:posterior_ex}. Details on how to control the appearance of this plot are provided in Section~\ref{subsec:plot_utils}.

\section{More Details}\label{sec:details}
In this section, we provide more details on the installation procedure (Section~\ref{subsec:install}), how to run \ptarcade{} (Section~\ref{sec:run}), how \texttt{PTArcade} implements user-specified signals in PTAs analysis pipelines (Section~\ref{subsec:code}), how to personalize a \texttt{PTArcade} run by using model and configuration files (Section~\ref{subsec:user_inputs}), the structure of the output data (Section~\ref{subsec:out}), and the content of the utility modules (Section~\ref{subsec:utils}). Throughout this section, user-controllable parameters will be highlighted in {\color{mygreen} green} and hyperlinked to the relevant discussion in the model or configuration file sections.

\subsection{Installation methods}
If you are familiar with Python and want to install \ptarcade{} on your local machine, we recommend doing so by using either \texttt{conda} or \href{https://pypi.org/project/pip/}{\texttt{pip}}. If you are not familiar with Python or want to install \ptarcade{} on a cluster, we recommend using a \href{https://www.docker.com/}{docker} or a \href{https://docs.sylabs.io/guides/2.6/user-guide/introduction.html}{singularity} container. In the following, we will discuss all these possible installation methods.
\subsubsection*{Conda installation {\scriptsize\textmd{recommended}}}
The easiest way of installing \texttt{PTArcade} is via the Python package manager \texttt{conda}. We are in the process of submitting \ptarcade{} to the conda forge-channel, and soon you will be able to install \ptarcade{} as a conda package (refer to the \href{https://andrea-mitridate.github.io/PTArcade/getting_started/local_install/}{"Installation section"} of the \ptarcade{} documentation web page for updates on this). In the meantime, you can install \ptarcade{} using conda by downloading \href{https://andrea-mitridate.github.io/PTArcade/assets/downloads/ptarcade.yml}{this environment file}, and typing in a terminal
\begin{lstlisting}[language=mybash,
numbers=left,
mathescape]
conda env create -f ptarcade.yml
\end{lstlisting}
This will install \ptarcade{} and all the required dependencies in a conda environment named \texttt{ptarcade}, and download the PTA data from the NANOGrav~\cite{McLaughlin:2013ira} and IPTA collaboration~\cite{Hobbs_2010}. If you do not have conda installed on your machine, you should follow the instructions at \href{https://conda.io/projects/conda/en/latest/user-guide/install/index.html}{this link}.

\subsubsection*{PyPI installation}\label{subsec:install}
\ptarcade{} is also released as a PyPI package and can be installed using\texttt{pip}. However, the \texttt{pip} installation requires that non-Python dependencies are installed separately on your machine (or in a virtual environment). These dependencies are \href{https://github.com/aarchiba/tempo2}{\texttt{tempo2}}, \href{https://github.com/DrTimothyAldenDavis/SuiteSparse}{\texttt{SuiteSparse}}, and an \href{https://www.mpi-forum.org/}{\texttt{MPI}} implementation. You can install these packages by typing the following in a terminal
\begin{itemize}
\item \texttt{tempo2}
\begin{lstlisting}[language=mybash,
numbers=left,
mathescape]
curl -sSL https://raw.githubusercontent.com/vallis/libstempo/master/install_tempo2.sh | /*@\color{draculapurple}sh@*/
\end{lstlisting}

\item  \texttt{suite-sparse}
\begin{lstlisting}[language=mybash,
numbers=left,
mathescape]
# on mac
brew install suite-sparse

# on debian
sudo apt install libsuitesparse-dev
\end{lstlisting}

\item \texttt{MPI}
\begin{lstlisting}[language=mybash,
numbers=left,
mathescape]
# on mac
brew install open-mpi

# on debian
sudo apt install libopenmpi-dev openmpi-bin
\end{lstlisting}

\end{itemize}
Once you have installed these dependencies, you can install \ptarcade{} from a terminal by typing
\begin{lstlisting}[language=mybash,
numbers=left,
mathescape]
pip install ptarcade
\end{lstlisting}
This will install all necessary Python dependencies and download the PTA data from the NANOGrav and IPTA collaboration.
\subsubsection*{With Docker\label{sec:install-docker}}
\ptarcade{} is also packaged as a Docker image---including its Python and non-Python dependencies.
You can download the image from a terminal by typing

\begin{lstlisting}[language=mybash,
numbers=left,
mathescape]
docker pull ngnewphy/ptarcade:latest
\end{lstlisting}

\subsubsection*{With singularity\label{sec:install-singularity}}
You create a Singularity image for \ptarcade{} by typing in your terminal

\begin{lstlisting}[language=mybash,
numbers=left,
mathescape]
singularity pull /*@\color{draculacl}ptarcade.sif docker://ngnewphy/ptarcade:latest@*/
\end{lstlisting}
This will create a Singularity image and save it as \texttt{ptarcade.sif} in the current working directory.
\subsection{Run \ptarcade{}}\label{sec:run}
If you have installed \ptarcade{} with \texttt{pip} or \texttt{conda}, you can run \ptarcade{} by opening a terminal and typing\footnote{Remember to activate the correct virtual environment if you installed \ptarcade{} inside one!}
\begin{lstlisting}[language=mybash,
numbers=left,
mathescape]
/*@\color{draculapurple}ptarcade@*/ -m ./model.py
\end{lstlisting}
Here, the argument passed to the \texttt{-m} input flag is the path to a model file (in this example we are passing a model file named \texttt{model.py} and located in the current directory). In addition to a model file, other two optional arguments can be passed when running \ptarcade{}:

\begin{itemize}
    \item A \emph{configuration file} can be passed via the input flag \texttt{-c}. Configuration files allow the user to control several aspect of the run, including the PTA dataset used, the number of MC trials, etc. More details on configuration files can be found in Section~\ref{subsec:config}.
    
    \item A string that will be appended to the MC chain folder. This can be useful if you are running multiple instances of \ptarcade{} for the same model (which can help you to get faster convergence) and you want to use the same output directory for all of them. By default, the chains will be saved in \texttt{./chains/np\_model/chain\_0}. Each of the three elements of this path can be controlled by the user. \texttt{./chains} can be changed by using the \hyperlink{config.out-dir}{{\color{mygreen}\texttt{out\_dir}}} parameter in the configuration file, \texttt{np\_model} can be changed by using the \hyperlink{model.name}{{\color{mygreen}\texttt{name}}} parameter in the model file, and \texttt{chain\_0} can be changed via the argument passed to the \texttt{-n} input flag of the \texttt{ptarcade} command. The argument passed to the \texttt{-n} flag will be appended to \texttt{chain\_}, so that passing \texttt{-n 42} will result in the output directory \texttt{./chains/np\_model/chain\_42}.
\end{itemize}

\subsection*{Run \ptarcade{} with Docker\label{sec:run-docker}}
The commands in Section~\ref{sec:run} must be slightly modified to run within a Docker container.
Docker does not mount any directories into the container by default. 
You must pass directories to mount inside the container using the syntax \texttt{-v <source>:<destination>}. 
In the example below, we assume that the only directories you will pass to the command line options of \ptarcade{} are accessible from your current working directory.

\begin{lstlisting}[language=mybash,
numbers=left,
]
docker run -v $(pwd):$(pwd) -w $(pwd) -i -t ptarcade -m ./model.py
\end{lstlisting}

\begin{itemize}
    \item \texttt{-v} tells Docker what to mount from the host computer and where to mount it in the container. Here, we mount the current working directory of the host into the container using its full path.
    \item \texttt{-w} sets the working directory of the container. In this case, it sets it to the current working directory that was just mounted.
    \item\texttt{-i -t} keeps \texttt{STDIN} open and allocates a pseudo-TTY 
\end{itemize}

The \texttt{ptarcade} in the \texttt{docker run} command refers to the name of the Docker image.
If you would like to run something else inside the container, then replace the \ptarcade{} options with the program to run.
For example, to run an interactive Bash shell

\begin{lstlisting}[language=mybash,
numbers=left,
]
docker run -v $(pwd):$(pwd) -w $(pwd) -i -t ptarcade bash
\end{lstlisting}

\subsection*{Run \ptarcade{} with Singularity}
As with Docker, the commands to run \ptarcade{} must be slightly modified to run using Singularity.
However, the commands are much simpler because Singularity will automatically mount your home directory inside the container.
Using the \texttt{ptarcade.sif} file you created in Section~\ref{sec:install-singularity}, type into a terminal

\begin{lstlisting}[language=mybash,
numbers=left,
]
singularity run ptarcade.sif -m ./model.py
\end{lstlisting}
You can also pass another command to run. 
For example, to start a Jupyter notebook type

\begin{lstlisting}[language=mybash,
numbers=left,
]
singularity run ptarcade.sif jupyter notebook
\end{lstlisting}

If you want an interactive shell, run the following command
\begin{lstlisting}[language=mybash,
numbers=left,
]
singularity shell ptarcade.sif
\end{lstlisting}

\subsection{Statistical tools and their implementation}\label{subsec:code}
In this section, we provide more details on the inner workings of \ptarcade{} and the implementation of the user-specified input in \texttt{ENTERPRISE} or \texttt{ceffyl}.

\subsubsection*{The PTA likelihood}
Searches for stochastic or deterministic signals with PTAs utilize the pulsars' timing residuals,  \pref{config.pta-data}{$\vec{\delta t}$}{pta-data}, which measure the discrepancy between the observed times of arrival (TOAs) of the pulses and the TOAs predicted by the pulsar timing model~\citep{Ramani:2020hdo, Lee:2020wfn, Lee:2021zqw}. Timing residuals receive contributions by any effect not captured in the timing models used to derive them. This includes not only instrumental and spin noise but also GWB signals and possible deterministic signals.
Specifically, we model the timing residuals as the sum of white noise, red noise, and small errors in the
fit to the timing-ephemeris parameter~\cite{NANOGrav:2020tig}:
\begin{equation}
\label{eq:res_model}
\vec{\delta t}=\vec{n}+\mat{F}\,\vec{a}+\mat{M}\,\vec{\epsilon}\,.
\end{equation}

The first term on the right-hand side of Eq.~\eqref{eq:res_model}, $\vec{n}$, describes the white noise that is assumed to be left in each of the $N_{\textrm{TOA}}$ timing residuals after subtracting all known systematics. White noise is assumed to be a zero-mean normal random variable, fully characterized by its covariance. Following standard conventions \cite{NANOGrav:2015aud, NANOGRAV:2018hou}, \ptarcade{} sets the parameters of this $N_{\textrm{TOA}}\times N_{\textrm{TOA}}$ covariance matrix to their maximum-posterior values as recovered from single-pulsar noise studies (see reference~\cite{NANOGrav:2020gpb, ng12_gwb} for NG12, \cite{aaa+23_detchar} for NG15, and \cite{Antoniadis:2022pcn} for IPTA DR2). 

The second term on the right-hand side of Eq.~\eqref{eq:res_model} describes time-correlated stochastic processes, including pulsar-intrinsic red noise and GWB signals. These processes are modeled using a Fourier basis of frequencies $i/T_{\textrm{obs}}$, where $i$ indexes the harmonics of the basis and $T_{\textrm{obs}}$ is the timing baseline, extending from the first to the last recorded TOA in the complete PTA data set. Since we are generally interested in processes that exhibit long timescale correlations, the expansion is truncated after \pref{config.red-components}{$N_{f,\textrm{red}}$}{red-components} frequency bins for the intrinsic red-noise component, and \pref{config.gwb-components}{$N_{f,\scriptscriptstyle{\textrm{GWB}}}$}{gwb-components} frequency bins for the GWB component.
This set of $N_f$ sine--cosine pairs evaluated at the different observation times are contained in the Fourier design matrix, $\mat{F}$. The Fourier coefficients of this expansion, $\vec{a}$, are assumed to be normally-distributed random variables with zero mean and the covariance matrix, $\langle\vec{a}\vec{a}^{\textrm T}\rangle=\mat{\phi}$, given by\footnote{For the case of IPTA DR2 data, dispersion measure variations are also modeled as a time-correlated red noise process.}
\begin{equation}
\label{eq:red_cov}
[\phi]_{(ak)(bj)}=\delta_{ij}\left(\Gamma_{ab}\Phi_{i}+\delta_{ab}\varphi_{a,i}\right).
\end{equation}
Here, $a$ and $b$ index the pulsars, $i$ and $j$ index the frequency harmonics, and \pref{config.corr}{$\Gamma_{ab}$}{corr} is the GWB overlap reduction function, which describes average correlations between pulsars $a$ and $b$ as a function of their angular separation in the sky. For an isotropic and unpolarized GWB, $\Gamma_{ab}$ is given by the Hellings \& Downs correlation~\citep{1983ApJ...265L..39H}, also known as “quadrupolar” or “HD” correlation.

The first term on the right-hand side of Eq.~\eqref{eq:red_cov} parameterizes the contribution to the timing residuals induced by a GWB in terms of the model-dependent coefficients $\Phi_i$. These coefficients can be related to the GWB energy density per logarithmic frequency interval, $d\rho_{\scriptscriptstyle\textrm{GW}}/d\ln f$, as a fraction of the closure density, $\rho_c$, via~\citep{Allen:1997ad}
\begin{equation}
\label{eq:OGW}
h^2\Omega_{\scriptscriptstyle\textrm{GW}}(f) \equiv \frac{h^2}{\rho_c}\frac{d\rho_{\scriptscriptstyle\textrm{GW}}(f)}{d\ln f}=\frac{8 \pi^4 f^5}{H_0^2/h^2}\,\frac{\Phi(f)}{\Delta f}\,.
\end{equation}
Here, $\Delta f=1/T_{\textrm{obs}}$ is the width of the $N_f$ frequency bins. $H_0$ is the present-day value of the Hubble rate, and $h$ is the reduced Hubble constant, $H_0 = h \times 100\,\textrm{km}\,\textrm{s}^{-1}\,\textrm{Mpc}^{-1}$. Finally, $\Phi(f)$ determines the coefficients $\Phi_i$ in Eq.~\eqref{eq:red_cov}, i.e., $\Phi_i = \Phi\left(i/T_{\textrm{obs}}\right)$. \ptarcade{} will build 
$h^2\Omega_{\scriptscriptstyle\textrm{GW}}$ from any \pref{model.spectrum}{\texttt{spectrum}}{spectrum} function defined by the user in the model file, and, if \pref{model.smbhb}{\texttt{smbhb}}{smbhb}\texttt{=True}, add to it the expected signal produced by SMBHBs. The latter is modeled as a power law of the form
\begin{align}
    h^2\Omega_{\scriptscriptstyle\textrm{GW}}(f) = \frac{2 \pi^2A_{\scriptscriptstyle\textrm{BHB}}^2}{3 H_0^2} \left(\frac{f}{\textrm{year}^{-1}}\right)^{5-\gamma_{\scriptscriptstyle\textrm{BHB}}}\textrm{year}^{-2}\,.
\end{align}
For a population of binaries whose orbital evolution is driven purely by GW emission, the expected spectral index is $\gbhb = 13/3$~\citep{Phinney-2001}. However, current observations and numerical simulations only provide weak constraints on the value of $\abhb$. A commonly adopted choice in the literature is a \pref{config.gamma-bhb}{constant value}{gamma-bhb} prior for $\gbhb$, and a (somewhat arbitrary) uniform prior, $[\pref{config.a-bhb-logmin}{\log_{10}A_{\scriptscriptstyle\textrm{BHB}}^{\textrm{min}}}{a-bhb-logmin},\pref{config.a-bhb-logmax}{\log_{10}A_{\scriptscriptstyle\textrm{BHB}}^{\textrm{max}}}{a-bhb-logmax}]$, for $\log_{10}\abhb$. A more sophisticated prior choice that connects the priors to the underlying SMBHB model has been proposed in \cite{aaa+23_newphys}. The authors of this work chose a 2D Gaussian prior for the SMBHB parameters, which was fitted to the distribution of $\abhb$ and $\gbhb$ derived by performing a power-law fit to the SMBHB populations simulated in \cite{aaa+23_smbh}. These Gaussian priors are available in \ptarcade{} and can be used by setting \pref{config.bhb-th-prior}{\texttt{bhb\_th\_prior}}{bhb-th-prior}\texttt{=True} in the configuration file.

The last term in Eq.~\eqref{eq:red_cov} models pulsar-intrinsic red noise in terms of the coefficients $\varphi_{a,i}$, where
\begin{equation}
\label{eq:kappaa}
\varphi_a(f) = \frac{A_a^2}{12\pi^2}\frac{1}{T_{\textrm{obs}}}\left(\frac{f}{\textrm{year}^{-1}}\right)^{-\gamma_a}\,\textrm{year}^3 
\end{equation}
and $\varphi_{a,i} = \varphi_a(i/T_{\textrm{obs}})$ for all $N_f$ frequencies. The priors (one per pulsar) for the amplitudes of the intrinsic red-noise processes are taken to be log-uniform in the range $[-20, -11]$, while the priors for the spectral indices are taken to be uniform in the range $[0,7]$. 

Finally, $\mat{M}\vec{\epsilon}$ accounts for deviations from the initial best-fit values of the $m$ timing model parameters. The \emph{design matrix}, $\mat{M}$, is an $N_{\textrm{TOA}}\times m$ matrix containing the partial derivatives of the TOAs with respect to each timing-ephemeris parameter evaluated at the initial best-fit value. $\vec{\epsilon}$ is a vector containing the linear offset from these best-fit parameters.

Since we are not interested in the specific realization of the noise but only in its statistical properties, we can analytically marginalize over all the possible noise realizations, i.e., integrate over all the possible values of $\vec{a}$ and $\vec{\epsilon}$. We are then left with a marginalized likelihood that depends only on the (unknown) parameters describing the red-noise covariance matrix. We collectively denote these parameters with $\vec\theta$, which includes $A_a$, $\gamma_a$, as well as any other parameters describing the user-specified signal. The likelihood reads~\citep{van_Haasteren_2012,Lentati_2013}: 
\begin{equation}
\label{eq:likelihood}
p(\vec{\delta t}|\vec\theta)=\frac{\exp\left(-\frac{1}{2}\vec{\delta t}^{ T}\mat{C}^{-1}\vec{\delta t}\right)}{\sqrt{{\textrm{det}}(2\pi\mat{C})}}\,,
\end{equation}
where $\mat{C}=\mat{N}+\mat{TBT}^{ T}$. Here, $\mat{N}$ is the covariance matrix of the white noise, $\mat{T}=[\mat{M},\mat{F}]$.  $\mat{B}=\textrm{diag}(\mat{\infty},\mat{\phi})$ where $\mat{\infty}$ is a diagonal matrix of infinities, which effectively means that we assume flat priors for the parameters in $\vec{\epsilon}$. Since in our calculations, we always deal with the inverse of $\mat{B}$, all these infinities reduce to zeros. If the user specifies a deterministic signal via the \preff{model.singal}{\texttt{signal}}{signal} function in the model file, Eq.~\eqref{eq:likelihood} will be modified by shifting the timing residuals as $\vec{\delta t}\to\vec{\delta t}-\vec{h}$, where $\vec{h}$ is an array containing the value of the \preff{model.singal}{\texttt{signal}}{signal} function evaluated at each of the TOAs.

Evaluating the likelihood given in Eq.~\eqref{eq:likelihood} requires inverting the covariance matrix, $\mat{C}$. When spatial correlations are included in the calculation (\hyperlink{config.corr}{{\color{mygreen}\texttt{corr}}}\texttt{=True}), $\mat{C}$ is a dense $N_{\textrm{TOA}}\times N_{\textrm{TOA}}$ matrix, and likelihood evaluation takes $\mathcal{O}(0.2\;\textrm{s})$~\cite{lamb2023need}. When spatial correlations are ignored (\hyperlink{config.corr}{{\color{mygreen}\texttt{corr}}}\texttt{=False}), the covariance matrix is block diagonal, and the likelihood evaluation time reduces to $\mathcal{O}(0.01\;\textrm{s})$~\cite{lamb2023need}. 
However, if we are interested in analyzing a stochastic signal, we can further improve upon this. We can follow the procedure outlined in~\cite{lamb2023need} with a subsequent fit of the stochastic-signal spectrum to the \emph{free spectrum} of the PTA data, i.e., a violin plot. The free spectrum effectively gives the posterior distribution of $\Phi_i$ at each sampling frequency: $p(\Phi_i|\vec{\delta t})$. Refitting a stochastic signal to the free spectrum can be effectively accomplished by using, instead of Eq.~\eqref{eq:likelihood}, the following PTA likelihood~\cite{lamb2023need}:
\begin{align}\label{eq:ceffyl_likelihood}
    p(\vec{\delta t}|\vec\theta) = \prod_{k=1}^{N_{f,\textrm{GWB}}}\left.\frac{p(\Phi_k|\vec{\delta t})}{p(\Phi_k)}\right|_{\Phi_k=\Phi_{\textrm{GWB}}(k/T_{\textrm{obs}};\,\vec\theta)}\,.
\end{align}
Here, $p(\Phi_k)$ is the prior probability of $\Phi_k$ adopted in the analysis used to derive the free-spectrum $p(\Phi_k|\vec{\delta t})$, and $\Phi_{\textrm{GWB}}(f;\,\vec\theta)$ is the GWB spectrum with $\vec{\theta}$ being the parameter of the model. 

\subsubsection*{Bayesian inference and MCMC sampler}
All the techniques implemented in \ptarcade{} use Bayesian inference to derive information on the parameters of the user-specified signal from the pulsars' timing residuals. Timing residuals measure the discrepancy between the observed pulse times of arrival and the ones predicted by the pulsar timing model (for more details on the timing model used in \ptarcade{} see e.g~\cite{aaa+23_timing}). Specifically, given the timing residuals, \pref{config.pta-data}{$\vec{\delta t}$}{pta-data}, and a set of parameters, $\vec\theta$, for the model that we use to describe them, we can use Bayes' theorem to write 
\begin{align}\label{eq:posterior}
    p(\vec\theta|\vec{\delta t}) = \frac{p(\vec{\delta t}|\vec\theta)p(\vec\theta)}{p(\vec{\delta t})}.
\end{align}
Here, $p(\vec{\theta}|\vec{\delta t})$ is the posterior probability distribution for the model parameters, $p(\vec{\delta t}|\vec{\theta})$ is the PTA likelihood, $p(\vec{\theta})$ is the prior probability distribution, and
\begin{equation}\label{eq:evidence}
    \mathcal{Z}\equiv p(\vec{\delta t})=\int \textrm{d}\vec{\theta}\: p(\vec{\delta t}|\vec{\theta})p(\vec{\theta})
\end{equation}
is the marginalized likelihood or evidence. We want to derive the posterior distribution, as it encodes the probability distribution for the model parameters (which include the parameters of the user-specified signal) given the observed data. 
If \pref{config.mode}{\texttt{mode}}{mode}\texttt{="enterprise"}, \ptarcade{} will use the PTA likelihood given in Eq.~\eqref{eq:likelihood} and implement it using \texttt{ENTERPRISE}~\cite{2019ascl.soft12015E} and \texttt{ENTERPRISE\_EXTENSIONS}~\cite{enterprise}. If \hyperlink{config.mode}{\color{mygreen}\texttt{mode}}\texttt{="ceffyl"}, \ptarcade{} will use the PTA likelihood given in Eq.~\eqref{eq:ceffyl_likelihood} and implement it with \texttt{ceffyl}~\cite{lamb2023need}.

While, in principle, the likelihood in Eq.~\eqref{eq:likelihood} is all we need to derive the (with respect to the noise and DM parameters) marginalized posteriors for the user-specified parameters. However, this is computationally expensive given the large dimensionality of the typical parameter space. Therefore, these integrals are performed by using Monte Carlo sampling. Specifically, \ptarcade{} uses the Markov chain Monte Carlo (MCMC) tools implemented in the \texttt{PTMCMCSampler} package~\citep{justin_ellis_2017_1037579} to sample \pref{config.n-samples}{$N_{\textrm{sam}}$}{n-samples} parameter points from the posterior distribution.

\subsubsection*{Model comparison}
\ptarcade{} can also be used to perform a model selection analysis between the user-specified model, $\mathcal{H}_1$, and the reference model, $\mathcal{H}_0$, where the only source for the GWB is provided by SMBHBs. Specifically, \ptarcade{} can be used to derive the Bayes factor defined as 
\begin{equation}\label{eq:Bayes Factor}
\mathcal{B}_{10}= \frac{\mathcal{Z}_1}{\mathcal{Z}_0} = \frac{p(\vec{\delta t}|\mathcal{H}_1)}{p(\vec{\delta t}|\mathcal{H}_0)} \,.
\end{equation}
To compute the Bayes factor between the two models, \ptarcade{} uses product space methods~\citep{10.2307/2346151, 10.2307/1391010, 10.1093/mnras/stv2217}. Correspondingly, if \pref{config.mod-sel}{\texttt{mod\_sel}}{mod-sel}\texttt{=True} in the configuration file, a model indexing variable controlling which model likelihood is active at each MCMC iteration will be sampled along with the parameters of the competing models. Then, the Bayes factor between models can be derived by taking the ratio of samples in each bin of the model indexing variable. The uncertainty on the Bayes factors obtained in this way can be derived by using statistical bootstrapping~\citep{Efron:1986hys}. When bootstrapping, new sets of Monte Carlo draws are created by resampling the original set of draws. We can then obtain a distribution for the Bayes factors from these independent realizations of the sampling procedure and compute the mean and the standard deviation of this distribution that we use to estimate the Bayes factor and its uncertainty. \ptarcade{} provides the function \preff{utils.compute-bf}{\texttt{compute\_bf}}{compute-bf} in the \texttt{chains\_utils} module to compute Bayes factors. This function can either compute the Bayes factor directly from the chain, as described below \eqref{eq:Bayes Factor}, or use the bootstrapping method.

\subsection{Output details}\label{subsec:out}
The output generated by \ptarcade{} matches that produced by \texttt{ENTERPRISE}, and it includes, beyond the MC chains, several files that summarize valuable information on the run and the MC sampler. By default, the structure of the output is the following:
\vspace{5pt}
\dirtree{%
.1 ./chains/.
.2 np\_model/.
.3 chain\_0/.
.4 chain\_1.txt.
.4 pars.txt.
.4 priors.txt.
.4 $\ldots$.
}
\vspace{5pt}
\noindent By default, the root directory for the output material is \texttt{./chains}. The user can change this using the \pref{config.out-dir}{\texttt{out\_dir}}{out-dir} parameter in the configuration file. Inside this root directory, the results of the current run are saved in a folder that, by default, is called \texttt{np\_model}. The user can change the name of this folder via the \pref{model.name}{\texttt{name}}{name} parameter in the model file. Finally, inside this folder, there will be one (or more, depending on how many chains you ran for this model) folder named \texttt{chain\_0}. In case you want to run multiple chains for the same model, it can be useful to store all your results in the same folder, as discussed in Section~\ref{sec:run}. You can do this by changing the number appended to the \texttt{chain\_} folder via the \texttt{-n} input flag when running \ptarcade{}.
For our purposes here, the most important files produced by PTArcade are:
 \begin{itemize}
     \item \preff{pars}{\texttt{pars.txt}}{out.pars}

    This file contains the names of the model parameters. The order in 
    which the parameters appear in this file will also dictate the order 
    in which the parameters appear in the \texttt{chain\_1.txt} file.

    When running with \hyperlink{config.mode}{\color{mygreen}\texttt{mode}}\texttt{="ceffyl"}, the \texttt{pars.txt} file for the example
    model discussed in Section~\ref{sec:quick} will read as follows: 

\begin{lstlisting}[language=mybash,
numbers=left,
mathescape]
log_A_star
log_f_star
\end{lstlisting}
    
    When running with \hyperlink{config.mode}{\color{mygreen}\texttt{mode}}\texttt{="enterprise"}, in addition
    to the user-specified parameters, \texttt{pars.txt} will also include intrinsic
    red noise parameters (two per pulsar) and, in the case of the IPTA dataset,
    DM parameters. 

    \item \preff{chain}{\texttt{chain\_1.txt}}{out.chain}

    This file contains the MC chains. It is formatted such that each
    line represents an MC sample, and each column corresponds to a 
    parameter of our model. The ordering of the parameters, i.e., which 
    column is associated with each parameter, can be read out from the 
    \preff{out.pars}{\texttt{pars.txt}}{pars} file. In addition to the model parameters, the last four columns of each row report the values of the posterior, the likelihood, the acceptance rate, and an indicator variable for parallel tempering, which does not matter in our case since \ptarcade{} does not use parallel tempering at the moment. For the example model discussed in Section~\ref{sec:quick}, the output of a run in Ceffyl-mode will be: 

\begin{lstlisting}[language=mybash,
numbers=left,
mathescape]
-7.893	-7.353	-15.606	4.715	-64.582	-60.960	0.507	1.0
-8.187	-7.509	-15.606	4.675	-65.498	-61.872	0.507	1.0
-8.088	-7.638	-15.741	4.674	-65.598	-61.943	0.507	1.0
\end{lstlisting}

    Here, the first two columns give the values of $\log_{10}A_*$ and $\log_{10}(f_*/{\textrm Hz})$ and the remaining columns give the value of the posterior, the likelihood, the acceptance rate, and the parallel-tempering indicator.
    Note that, when running with \hyperlink{config.mode}{\color{mygreen}\texttt{mode}}\texttt{="enterprise"}, in addition to the user-specified parameters, the chains will also include intrinsic red noise parameters (two per pulsar) and, in the case of the IPTA dataset, DM parameters. 

    \item\preff{priors}{\texttt{priors\_1.txt}}{out.priors}

    The prior file is similar to the \texttt{priors.txt} file, but it includes their prior distributions in addition to the parameter names. Here is an example of our test model when running in Ceffyl mode.

\begin{lstlisting}[language=mybash,
numbers=left,
mathescape]
log_A_star:Uniform(pmin=-14, pmax=-6)
log_f_star:Uniform(pmin=-10, pmax=-6)
\end{lstlisting}
 \end{itemize}

\subsection{User inputs}\label{subsec:user_inputs}
When running \ptarcade{}, the user can provide two input files:

\begin{itemize}
    \item[--] Model file · {\bf Required} – This file, passed via the \texttt{-m} input flag, contains the definition of the new-physics signal. In the case of stochastic signals, this boils down to defining the GWB energy density per logarithmic frequency interval. In the case of deterministic signals, the user should define the time series of induced timing delays, $h(t)$, in units of seconds.

    \item[--] Configuration file · {\bf Optional} – In addition to the model file, the user can pass a configuration file via the input flag \texttt{-c}. The configuration file is a simple Python file that allows the user to adjust several run parameters.
\end{itemize}

In this section, we will discuss in detail the structure and the functionalities of these two input files.

\subsubsection*{Model file}
The model file is a simple Python file that allows the user to define their model. At minimum, the model file needs to contain the two following information: 
\begin{itemize}
    \item \pref{parameters}{\texttt{parameters}}{model.parameters}

    {\small This variable needs to be assigned to a dictionary. The keys of this dictionary must be strings, which will be used as names for the model parameters. The values of this dictionary are \href{https://enterprise.readthedocs.io/en/latest/enterprise.signals.html#enterprise.signals.parameter.Parameter}{\texttt{ENTERPRISE} Parameter} objects. The user can create these objects via the \texttt{prior} helper function defined in the \texttt{models\_utils} module. The first argument passed to the \texttt{prior} function needs to be a string identifying the prior type. The following arguments are the parameters of the selected prior type. Several of the most common priors are already implemented in \ptarcade. They are listed in Table~\ref{tab:priors_tab} with their associated parameters and functional form. By default, the parameters are assumed to be common across pulsars. If you want to specify a pulsar-dependent parameter, you can pass \texttt{common=False} as a keyword argument to the \texttt{prior} function.}
\end{itemize}
\begin{multicols}{2}
    For stochastic signals
    \begin{itemize}
    \item \pref{spectrum}{\texttt{spectrum}}{model.spectrum}

    {\small Stochastic signals are defined via the \texttt{spectrum} function. The first parameter of this function should be named \texttt{f}, and it is supposed to be a NumPy array containing the frequencies (in units of Hz) at which to evaluate the spectrum. The names of the remaining parameters should match the keys of the \texttt{parameters} dictionary. The \texttt{spectrum} function should return a NumPy array containing the value of $h^2\Omega_{\mathrm{GW}}$ at each of the frequencies in \texttt{f}.}
    \end{itemize}
    \columnbreak
    For deterministic signals\footnote{Note that \hyperlink{conig.mode}{\color{mygreen}\texttt{mode}}\texttt{="enterprise"} is required to analyze deterministic signals.}
    \begin{itemize}
    \item \pref{signal}{\texttt{signal}}{model.signal}:

    {\small Deterministic signals are defined via the \texttt{signal} function. The first parameter of this function should be named \texttt{toas} and it is supposed to be a NumPy array containing the times of arrival (TOAs) (in units of seconds) at which to evaluate the deterministic signal. The name of the remaining parameters should match the keys of the \texttt{parameters} dictionary. The \texttt{signal} function should return a NumPy array with the same dimensions as \texttt{toas} containing the value of the induced shift for each TOA in \texttt{toas}.}
    \end{itemize}
\end{multicols}

\begin{table}[]
\centering
\small
\renewcommand{\arraystretch}{2}
\setlength\tabcolsep{5pt}
\begin{tabular}{cccc}
Prior        & Functional form                          & Identifier   & Parameters \\ \hline
Uniform      & $(\texttt{pmax}-\texttt{pmin})^{-1}$     & \texttt{"Uniform"} & (\texttt{pmin}, \texttt{pmax})      \\
Normal       & $\dfrac{(2\pi)^{-\texttt{size}/2}}{\sqrt{\textrm{det}(\texttt{sigma})}}\exp\left(-\frac{1}{2}(\boldsymbol{x}-{\bf \texttt{mu}})^{\textrm T}\texttt{cov}^{-1}(\boldsymbol{x}-{\bf \texttt{mu}})\right)$                                      & \texttt{"Normal"}  & (\texttt{mu}, \texttt{cov}, \texttt{size})   \\
Exponential  &                  $\text{if:  }\texttt{pmin}\leq x \leq \texttt{pmax}: \ln(10)\cdot \frac{10^x}{10^{\texttt{pmax}}-10^{\texttt{pmin}}}; \;\text{else: } 0$                        & \texttt{"LinearExp"} & (\texttt{pmin}, \texttt{pmax}) \\
Constant  &  $\delta(x - \texttt{val})$                                        & \texttt{"Constant"} &   (\texttt{val}) \\
Gamma  &   $\frac{1}{\texttt{scale}\cdot \Gamma(a)}\left(\frac{x-\texttt{loc}}{\texttt{scale}}\right)^{a-1}\exp\left(-\frac{x-\texttt{loc}}{\texttt{scale}}\right)$                                       & \texttt{"Gamma"} &  (\texttt{a}, \texttt{loc}, \texttt{scale})\\\hline
\end{tabular}
\caption{Common prior functions already implemented in \ptarcade{}. The second column gives the expression for the distribution, the third column gives the string that needs to be passed to the \texttt{prior} function to select the corresponding prior, and the last column gives the prior parameters that need to be passed to the \texttt{prior} function in addition to the string identifier.\label{tab:priors_tab}}
\end{table}

The model file can also contain additional (optional) variables that control the new-physics signal in more detail. To be specific, you can control the following:

\begin{itemize}
    \item \pref{name}{\texttt{name}}{model.name}:

    {\small {\bf Default: \texttt{"np\_model"}} -- This variable can be assigned to a string to specify the model name. Its value determines the name of the output directory associated with the model file.}
    
    \item \pref{smbhb}{\texttt{smbhb}}{model.smbhb}:

     {\small {\bf Default: \texttt{False}} -- If set to \texttt{True}, the expected signal from SMBHBs will be added to the user-specified signal.}
\end{itemize}

The model files used in the NANOGrav 15-year search for new-physics \cite{aaa+23_newphys} can be found \href{https://zenodo.org/record/8084351}{here}.

\subsubsection*{Configuration file}\label{subsec:config}
The configuration file is a Python file that allows the user to adjust several 
run parameters. The parameters that can be set in the configuration file are:

\begin{itemize}
    \item \pref{pta-data}{\texttt{pta\_data}}{config.pta-data}
    
    {\small {\bf Default: \texttt{"NG15"}} -- This variable needs to be assigned to a string specifying the PTA dataset to be used in the analysis. The datasets currently implemented in PTArcade are NANOGrav 15-year (\texttt{pta\_data="NG15"})~\cite{aaa+23_timing}, NANOGrav 12.5-year (\texttt{pta\_data="NG12"})~\cite{NANOGrav:2020gpb}, and IPTA DR2 (\texttt{pta\_data="IPTA2"})~\cite{Antoniadis:2022pcn}}.
    
    \item \pref{n-samples}{\texttt{N\_samples}}{config.n-samples}:

    {\small {\bf Default: \texttt{2e6}} -- This variable can be assigned to an integer, which specifies the number of points generated by the  Monte Carlo sampler. Note that the MC chains are automatically thinned by a factor of 10 to reduce the auto-correlation length. Therefore, the number of MC samples that will be saved is given by \texttt{N\_samples}$/10$.}
    
    \item \pref{mode}{\texttt{mode}}{config.mode}:

    {\small {\bf Default: \texttt{ceffyl}} -- \ptarcade{} can be run in two modes:

    \begin{itemize}
        \item \texttt{mode="enterprise"}: In this configuration, the code will analyze the PTA dataset at the level of the timing residuals and use the PTA likelihood given in Eq.~\eqref{eq:likelihood}.
        \item \texttt{mode="ceffyl"}: In this configuration, the code will analyze the PTA dataset at the level of the Bayesian periodograms and use the PTA likelihood given in Eq.~\eqref{eq:ceffyl_likelihood}.
    \end{itemize}}

    \item \pref{out-dir}{\texttt{out\_dir}}{config.out-dir}:

    {\small {\bf Default: \texttt{"./chains"}} -- This variable can be assigned to a string to specify the output directory.}
    
    \item \pref{resume}{\texttt{resume}}{config.resume}:

    {\small {\bf Default: \texttt{"./chains"}} If \texttt{resume=True}, the code will look for MCMC chains in the output directory and, if it finds any, it will restart sampling from those instead of starting from scratch. If \texttt{resume = True}, but there are no existing chains in the output directory, the sampler will start from scratch.}
        
    \item \pref{mod-sel}{\texttt{mod\_sel}}{config.mod-sel}:

    {\small {\bf Default: \texttt{False}} --  If \texttt{mod\_sel=True}, a model-indexing variable controlling which model likelihood is active at each MCMC iteration will be sampled along with the parameters of the competing model. This setup will then allow one to derive the Bayes factor between models by simply taking the ratio of samples spent in each bin of the model-indexing variable (see Section~\ref{subsec:chain_utils} for more details on this \ptarcade{} feature). Notice that, at the moment, model selection is only available if \hyperlink{config.mode}{\color{mygreen}\texttt{mode}}\texttt{="enterprise"}.}
    
    \item \pref{corr}{\texttt{corr}}{config.corr}:

    {\small {\bf Default: \texttt{"False"}} -- If set to \texttt{True}, the overlap reduction function for the common red noise term in Eq.~\eqref{eq:red_cov} will be set to the HD correlation; if set to \texttt{False}, the code will ignore cross-correlations and set $\Gamma_{ab}=\delta_{ab}$.}
    
    \item \pref{red-components}{\texttt{red\_components}}{config.red-components}

     {\small {\bf Default: \texttt{30}} -- This variable can be assigned to an integer specifying the number of frequency components that model the intrinsic red noise.}
    \item \pref{gwb-components}{\texttt{gwb\_components}}{config.gwb-components}

    {\small {\bf Default: \texttt{14}} -- This variable can be assigned to an integer specifying the number of frequency components that model the common red noise produced by a GWB.}
    
    \item \pref{bhb-th-prior}{\texttt{bhb\_th\_prior}}{config.bhb-th-prior}:

    {\small {\bf Default: \texttt{True}} -- If \texttt{bhb\_th\_prior=True}, the prior for the SMBHB signal parameters will be chosen to reflect predictions from astrophysical models. This is only relevant if you have selected \hyperlink{model.smbhb}{\color{mygreen}\texttt{smbhb}}\texttt{=True} in the model file or \hyperlink{config.mod-sel}{\color{mygreen}\texttt{mod\_sel}}\texttt{=True} in the configuration file.}
    
    \item \pref{a-bhb-logmin}{\texttt{A\_bhb\_logmin}}{a-bhb-logmin}

    {\small {\bf Default: \texttt{-18}} -- This variable can be assigned to a floating point or integer number to set the lower bound on the log-uniform prior of the SMBHB-signal amplitude. This is only relevant if \hyperlink{config.bhb-th-prior}{\color{mygreen}\texttt{bhb\_th\_prior}}\texttt{=False} and you have selected \hyperlink{model.smbhb}{\color{mygreen}\texttt{smbhb}}\texttt{=True} in the model file or \hyperlink{config.mod-sel}{\color{mygreen}\texttt{mod\_sel}}\texttt{=True} in the configuration file.}

    \item \pref{a-bhb-logmax}{\texttt{A\_bhb\_logmax}}{a-bhb-logmax}
    
    {\small {\bf Default: \texttt{-14}} -- This variable can be assigned to a floating point or integer number to set the upper bound on the log-uniform prior of the SMBHB-signal amplitude. This is only relevant if \hyperlink{config.bhb-th-prior}{\color{mygreen}\texttt{bhb\_th\_prior}}\texttt{=False} and you have selected \hyperlink{model.smbhb}{\color{mygreen}\texttt{smbhb}}\texttt{=True} in the model file or \hyperlink{config.mod-sel}{\color{mygreen}\texttt{mod\_sel}}\texttt{=True} in the configuration file.}
    
    \item \pref{gamma-bhb}{\texttt{gamma\_bhb}}{config.gamma-bhb}

    {\small {\bf Default: \texttt{None}} -- This variable can be assigned to a floating point or integer number to set the value of $\gamma_{\textrm{BHB}}$. If \texttt{gamma\_bhb=None}, a uniform prior between $0$ and $7$ will be used instead. This is only relevant if \hyperlink{config.bhb-th-prior}{\color{mygreen}\texttt{bhb\_th\_prior}}\texttt{=False} and you have selected \hyperlink{model.smbhb}{\color{mygreen}\texttt{smbhb}}\texttt{=True} in the model file or \hyperlink{config.mod-sel}{\color{mygreen}\texttt{mod\_sel}}\texttt{=True} in the configuration file.}
\end{itemize}

If the user passes no configuration file to the \texttt{-c} flag, the default configuration file is the following:
\begin{lstlisting}[language=mypython, 
title=Default configuration file,
numbers=left,
mathescape]
pta_data = 'NG15'

mode = 'ceffyl'

mod_sel = False

out_dir = './chains/'
resume = False 
N_samples = int(2e6) 

# intrinsic red noises parameters
red_components = 14 

# bhbh signal parameters
corr = False 
gwb_components = 14 
bhb_th_prior = True 
\end{lstlisting}

\subsection{Utilities}\label{subsec:utils}

\ptarcade{} comes with several handy utility modules, which are designed to assist the user in building model files, evaluating MCMC chains, and creating posterior plots.
\begin{itemize}
    \item[--] \texttt{ptarcade.models\_utils} -- This module aims at facilitating the creation of model files. It contains several useful constants expressed in natural units, a parametrization of the effective number of relativistic degrees of freedom contributing to the Universe's energy and entropy densities, $g_\rho$ and $g_s$, and a function to define spectra for stochastic signals from tabulated data.
    \item[--] \texttt{ptarcade.chains\_utils} -- This module serves two primary functions: Loading chains and model parameters of a \ptarcade{} run and computing the Bayes factor and its error for a \ptarcade{} run comparing, e.g., a GWB of new-physics origin to a GWB produced solely from SMBHBs.
    \item[--] \texttt{ptarcade.plot\_utils} -- This module contains functions to produce trace plots of MCMC chains to allow for visualization of their convergence. Moreover, it can be utilized to plot the 1D- and 2D-posteriors for all parameters of interest.
\end{itemize}
\subsubsection*{Model Utilities}
\begin{itemize}
    \item The constants and conversion factors reported in Table~\ref{tab:const} can be loaded from \texttt{models\_utils}. Unless otherwise specified, they are all expressed in natural units and their values are taken from \cite{ParticleDataGroup:2020ssz}.

\item \normalsize \texttt{models\_utils.g\_rho} \& \texttt{models\_utils.g\_s}

{\small These functions return the effective number of relativistic degrees of freedom contributing to the Universe's energy and entropy density at a given temperature $T$ (in GeV) or as a function of frequency $f$ (in Hz). In the latter case, they return the value of these functions at the time of the cosmological evolution when GWs with comoving wavenumber $k=2\pi a_0 f$ re-entered the horizon. Here, $a_0$ denotes the value of the cosmological scale factor today, which we set to $a_0=1$. The functions are derived by interpolating the tabulated data in \cite{Saikawa_2020}. In the following example, we evaluate $g_\rho(T)$ and $g_s(f)$ for $T=1\, \mathrm{GeV}$ and $f = \left(10^{-9}\, \mathrm{Hz},\, 10^{-8}\, \mathrm{Hz},\, 10^{-7} \, \mathrm{Hz}\right)$.
\begin{lstlisting}[language=mypython, 
numbers=left,
mathescape,
xleftmargin = 2.em]
import numpy as np
from ptarcade.models_utils import g_rho, g_s

T = 1. #(in GeV)
g_1GeV = g_rho(T)

f = np.array([1e-9, 1e-8, 1e-7]) #(in Hz)
g_sf = g_s(f, is_freq=True)
\end{lstlisting}

Note, that \texttt{g\_rho} and \texttt{g\_s} accept any array-like input in the first argument and will return an array-like object with the same dimensions. The second argument is a \texttt{bool}. By default, this boolean is set to \texttt{False}, indicating that the first argument is a temperature (in units of GeV). If it is set to \texttt{True}, the first argument is assumed to be a frequency (in units of Hz).}

\item \texttt{model\_utils.spec\_importer}

{\small This function allows to define the spectrum of a stochastic signal by using tabulated data. This is useful if the spectrum you are interested in is only evaluated numerically without a closed analytical expression for the stochastic signal amplitude $h^2 \Omega_{\textrm{GW}}$. \texttt{spec\_importer} expects the path to an HDF5 file containing the spectrum as a function of frequency and eventual other parameters. It returns a callable function of the frequency $f$ and any other relevant parameters. In the example below, we interpolate a spectrum parametrized by frequency and one additional parameter $p$.
\begin{lstlisting}[language=mypython, 
numbers=left,
mathescape]
import os
from ptarcade.models_utils import spec_importer

path = "/This/Is/A/Path/to/the/HDF5/File/spectrum.h5"

log_spectrum = spec_importer(path)

def spectrum(f, p):
    return 10**log_spectrum(np.log10(f), p = p)
\end{lstlisting}
In this example, the HDF5 file was generated from a plain-text file with the following formatting:
\begin{lstlisting}[language=mybash,
numbers=left,
mathescape]
p	 f	        spectrum
-1	 -10.000000	-19.000000
-1	  -9.950000	-18.900000
-1	  -9.900000	-18.800000
...
-0.9 -10.000000	-19.100000
-0.9  -9.950000	-19.000000
-0.9  -9.900000	-18.900000
...
\end{lstlisting}
\ptarcade{} provides \texttt{fast\_interpolate.reformat} to convert such plain-text files to an HDF5 file that \texttt{fast\_interpolate.interp} will use to quickly interpolate tabulated data.
The plain-text files must meet the following requirements:
\begin{itemize}
    \item The file has a header with at least \textbf{spectrum} and \textbf{f} present
    \item Each column is evenly spaced
    \item The \textbf{f} column must be last if \textbf{spectrum} is not. If \textbf{spectrum} is last, \textbf{f} must be the second-to-last column. 
\end{itemize}
\texttt{fast\_interpolate.reformat} will convert the supplied plain-text file to an HDF5 file at a specified destination with the the following HDF5 datasets:
\begin{itemize}
    \item \texttt{parameter\_names} - this dataset contains the parameter names from the header other than \textbf{spectrum} 
    \item \texttt{spectrum} - this dataset contains the \textbf{spectrum} data from the original file
    \item There will be one additional dataset for each parameter other than \textbf{spectrum}. These datasets will contain two values: the minimum value the parameter can take and the step size. The example file above would generate such datasets for \textbf{f} and \textbf{p}. Assuming the HDF5 file has been read into memory as \textbf{data}, then you would have the following:
\begin{lstlisting}[language=mypython, 
numbers=left,
mathescape]
print(data["p"])
[-1.0, 0.1]

print(data["f"]
[-10.0, 0.05]
\end{lstlisting}

\end{itemize}}
\end{itemize}

\subsubsection*{Chain Utilities}
\label{subsec:chain_utils}
\begin{itemize}
    \item \preff{import-chains}{\texttt{chains\_utils.import\_chains}}{utils.import-chains}
    
    {\small This function can be used to load chains and model parameters of a \ptarcade{} run. It expects only one argument, namely the path to a folder containing chains generated using \ptarcade{} (for the default output folder structure discussed in Section~\ref{subsec:out}, the user should pass the path to the \texttt{np\_model} folder). The function loads all chains within the specified folder, merges them, and returns the resulting merged chain as a NumPy array together with a dictionary containing the parameters of the run and their priors. By default, \texttt{import\_chains} removes 25\% of each chain before merging. You can change the amount of burn-in by using the \texttt{burn\_frac} argument, specifying the fraction of each chain that is to be discarded. Also, note that by default, \texttt{import\_chains} only loads the part of the chains corresponding to user-specified parameters, the likelihood, the posterior, and the hypermodel index. If you also want to load red noise and eventual DM parameters, you can do so by setting the flag \texttt{quick\_import=False}.}
    
    \item \preff{compute-bf}{\texttt{chains\_utils.compute\_bf}}{utils.compute-bf} 
    
    {\small This function can be used to compute Bayes factors from runs for which \hyperlink{config.mod-sel}{\color{mygreen}\texttt{mod\_sel}}\texttt{=True} is set in the configuration file (see Section~\ref{subsec:config}). The expected inputs are a chain and a parameter file in the output format of \hyperlink{utils.import-chains}{\texttt{import\_chains}}. The function returns an estimate for the Bayes factor comparing the user-specified signal against the SMBHB signal and the associated error. By default, the Bayes factor is calculated by dividing the number of points in the chain that fall in the hypermodel bin of the user-specified signal by the number of points falling in the bin of the reference SMBHB model. For a more precise estimate of the error on the Bayes factor, you can set \texttt{bootstrap=True}. In this case, the Bayes factor and its standard deviation will be derived by using bootstrapping methods.
    \begin{lstlisting}[language=mypython, 
numbers=left,
mathescape]
import ptarcade.chains_utils as utils

params, chain = utils.import_chains("path_to_chains_folder")

bf, bf_err = utils.compute_bf(chain, params)
\end{lstlisting}}
\end{itemize}

\subsubsection*{Plot Utilities}\label{subsec:plot_utils}
\begin{itemize}
    \item \preff{plot-chains}{\texttt{plot\_utils.plot\_chains}}{utils.plot-chains} 
    
    {\small This function produces trace plots of chains from \ptarcade{} runs. It expects a chain and the associated parameter dictionary in the output format of \preff{utils.import-chains}{\texttt{import\_chains}}{import-chains}. For example, Figure~\ref{fig:trace_ex} is produced from a chain stored in \texttt{'./chains/np\_model/'}. We can load the chain using \hyperlink{utils.import-chains}{\texttt{import\_chains}} and then pass the merged chains and parameters to \texttt{plot\_utils.plot\_chains}.
    \begin{lstlisting}[language=mypython, 
numbers=left,
mathescape]
from ptarcade import chains_utils as c_utils
from ptarcade import plot_utils as p_utils
        
params, chain = c_utils.import_chains('./chains/np_model/')
        
p_utils.plot_chains(chain, params)
    \end{lstlisting}
    Customization of the trace plots is possible using the optional arguments \texttt{params\_name} and \texttt{label\_size}. You can produce trace plots for a selected set of parameters or choose the desired format for the y-axis labels using \texttt{params\_name}. This argument expects  a dictionary containing the names of desired parameters as keys and the desired labels as values. By default, \texttt{params\_name=None}, in which case all pulsar-common parameters and all MCMC parameters are plotted without any additional formatting. The \texttt{label\_size} argument expects an \texttt{int}, by default \texttt{label\_size=13}, specifying the font size for the axis and tick labels.
    \begin{figure}[h]
        \centering
        \includegraphics[width=0.9\textwidth]{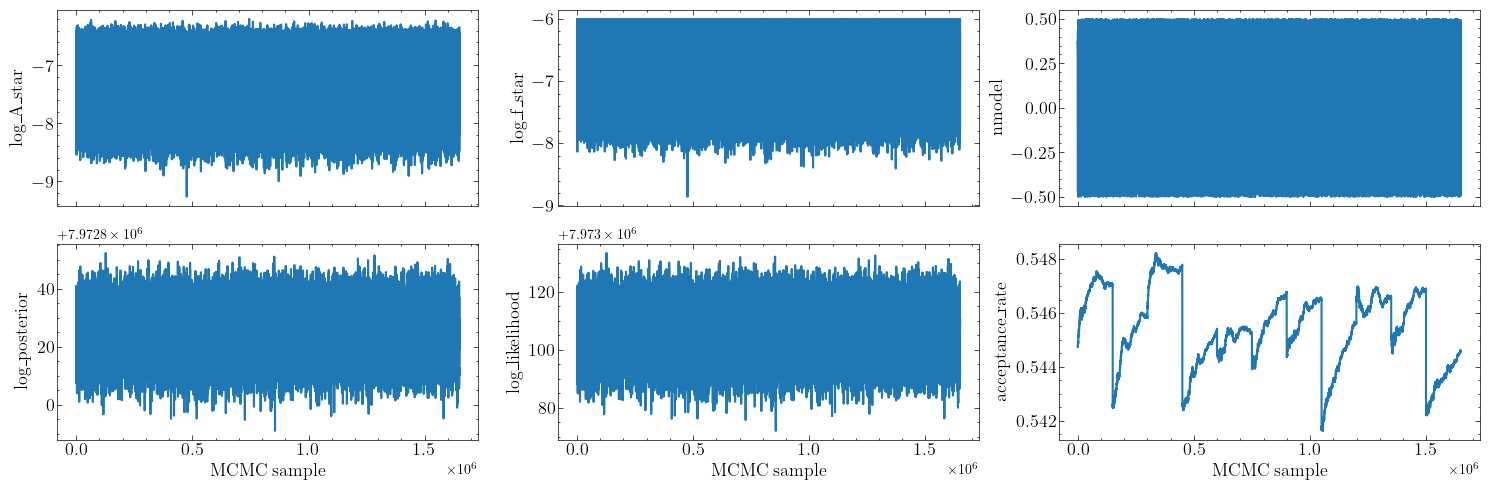}
        \caption{\small Trace plots for the default settings of \texttt{plot\_chains} for a model with two pulsar-common parameters, \texttt{"log\_A\_star"} and \texttt{"log\_f\_star"}, and a merged chain of length $\sim 1.7\times 10^6$ after thinning and removal of the cut-off.}
        \label{fig:trace_ex}
    \end{figure}}
    
    \item \preff{plot-posteriors}{\texttt{plot\_utils.plot\_posteriors}}{utils.plot-posteriors} 
    
    {\small This function produces posterior plots from MCMC chains, see for example Figure~\ref{fig:posterior_ex}. These plots are created using the \texttt{GetDist} package~\cite{Lewis:2019xzd}. This package provides kernel density estimation of the marginalized 1D- and 2D-posterior densities obtained from MC sampling to produce smooth posterior plots. Note that for \texttt{plot\_posteriors}, both the arguments \texttt{chains} and \texttt{params} are expected to be a list of chains and a list of parameter dictionaries, even if these lists contain only one item. You can superimpose the results of several runs by adding additional chains and parameter dictionaries to these lists. This is useful to compare a run set up with \hyperlink{model.smbhb}{\color{mygreen}\texttt{smbhb}}\texttt{=True} to a run without the SMBHB contribution. Several handy optional arguments for this function allow for user customization of the posterior plots. Here, we highlight the most useful features. For a more comprehensive list, please refer to the \href{https://andrea-mitridate.github.io/PTArcade/reference/ptarcade/plot_utils/}{"Reference section"} of \ptarcade{} documentation web page.
    \begin{itemize}
        \item As for \preff{utils.plot-chains}{\texttt{plot\_chains}}{plot-chains}, you can specify the parameters to be plotted. This is done by passing a list of lists containing the desired parameter names for each model to \texttt{par\_to\_plot}. By default, all pulsar-common parameters are plotted. 
        \item You can adjust axis labels by passing a list of lists containing the desired formats for each parameter to \texttt{par\_to\_plot}. By default, the labels are the parameter names in \texttt{params}.
        \item When analyzing a run with \hyperlink{config.mod-sel}{\color{mygreen}\texttt{mod\_sel}}\texttt{=True}, you can specify which hypermodel you want to plot using the \texttt{model\_id} argument. It expects a list of zeros and ones, which select the desired hypermodel for each chain in \texttt{chains}. The default value is \texttt{None}, in which case hypermodel zero is plotted for every chain.
        \item You can choose the confidence levels at which the highest posterior density intervals (HPI) are computed and shown in the 1D-posterior plots. The argument \texttt{hpi\_levels} expects a list of \texttt{float}s between zero and one corresponding to the desired confidence levels. The default, \texttt{hpi\_levels=[0.68, 0.95]}, corresponds to 68\% and 95\% confidence levels. Note that confidence levels are specified for every chain in \texttt{chains}, not on a chain-by-chain basis.
        \item You can choose the level at which to compute and plot the K-ratio bound using the \texttt{k\_ratio} argument. This argument expects a list of \texttt{float}'s between zero and one corresponding to the desired K-ratio levels which were introduced in \cite{aaa+23_newphys}. Each element in the list is associated with a chain in \texttt{chains}. The default value is \texttt{None}, in which case the K-ratio is not plotted. The K-ratio depends on the Bayes factor for the selected chain, and is, therefore, only sensible for \ptarcade{} runs with \hyperlink{config.mod-sel}{\color{mygreen}model selection}. For this feature to function properly, you need to pass a list of \hyperlink{utils.compute-bf}{Bayes factors} to the \texttt{bf} argument, where each element in \texttt{bf} was previously determined from the corresponding chain in \texttt{chains}. We keep the Bayes factors as an external input since computing the Bayes factor for a chain can be both time intensive and computationally expensive.
        \item The 2D-posterior plots generated by \hyperlink{utils.plot-posteriors}{\texttt{plot\_posteriors}} are shown as contour plots. You can adjust the confidence levels at which these contours are drawn using the \texttt{levels} argument. It expects a list of \texttt{float}s between zero and one, corresponding to the desired confidence level. The default value is \texttt{None}, in which case the contours correspond to 68\% and 95\% confidence level.
        \item By setting the \texttt{verbose} argument to \texttt{True}, you can print a statistical summary for the chains in \texttt{chains}. This contains information on the confidence intervals, the K-ratio, the highest posterior density points, and the Bayes estimator for the 1D-marginalized posterior distributions of every parameter specified to \hyperlink{utils.plot-posteriors}{\texttt{plot\_posteriors}}. The K-ratio and HPI intervals are determined based on the values passed to \texttt{k\_ratio} and \texttt{hpi\_values}.
    \end{itemize}}
        
\end{itemize}

\section{Acknowledgments}
We thank our colleagues in NANOGrav for fruitful discussions and feedback during the development of these tools. We particularly thank Luke Z. Kelley for helping derive the priors for the SMBHB signal, William Lamb for helping with the \texttt{ceffyl} implementation, and Rafael R. Lino dos Santos for helping debug \ptarcade{}. 
This work was supported by the Deutsche Forschungsgemeinschaft under Germany’s Excellence Strategy - EXC 2121 Quantum Universe - 390833306. \\
Part of this work was conducted using the High Performance Computing Cluster PALMA II at the University of M\"unster~(\url{https://www.uni-muenster.de/IT/HPC}). This work used the Maxwell computational resources operated at Deutsches Elektronen-Synchrotron DESY, Hamburg (Germany). This work was conducted in part using the HPC resources of the Texas Advanced Computing Center (TACC) at the University of Texas at Austin. The Tufts University High Performance Computing Cluster (\url{https://it.tufts.edu/high-performance-computing}) was utilized for some of the research reported in this paper. This research used the computational resources provided by the University of Central Florida's Advanced Research Computing Center (ARCC).\\
The work of R.v.E., K.Sc., and T.S.\ is supported by the Deutsche Forschungsgemeinschaft (DFG) through the Research Training Group, GRK 2149: Strong and Weak Interactions – from Hadrons to Dark Matter. 
A.Mi.\ is supported by the Deutsche Forschungsgemeinschaft under Germany's Excellence Strategy - EXC 2121 Quantum Universe - 390833306.
K.D.O.\ was supported in part by NSF grants Nos.\ 2111738 and 2207267.
T.T.\ contribution to this work is supported by the Fermi Research Alliance, LLC, under contract No. DE-AC02-07CH11359 with the U.S. Department of Energy, Office of Science, Office of High Energy Physics.

\begin{table}[h!]
\small{
\renewcommand{\arraystretch}{0.8}
\begin{center}
    \begin{tabular}{l l}
        \bf Attribute & \bf Description \\
        \hline
        & \\
        \texttt{G} & Newton's constant ($\mathrm{GeV}^{-2}$) \\
         & \textbf{TYPE:} \texttt{np.float64} \\
         & \\
         \texttt{M\_pl} & Reduced Planck mass (GeV) \\
         & \textbf{TYPE:} \texttt{np.float64} \\
          & \\
         \texttt{T\_0} & Present-day temperature of the Universe ($\mathrm{GeV}$) \\
         & \textbf{TYPE:} \texttt{np.float64} \\
          & \\
         \texttt{z\_eq} & Redshift of matter-radiation equality \\
         & \textbf{TYPE:} \texttt{int} \\
          & \\
         \texttt{T\_eq} & Temperature of matter-radiation equality (GeV) \\
         & \textbf{TYPE:} \texttt{np.float64} \\
          & \\
         \texttt{h} & Reduced Hubble parameter\\
         & \textbf{TYPE:} \texttt{float} \\
          & \\
         \texttt{H\_0} & Hubble constant (GeV) \\
         & \textbf{TYPE:} \texttt{np.float64} \\
          & \\
         \texttt{H\_0\_Hz} & Hubble constant (Hz) \\
         & \textbf{TYPE:} \texttt{np.float64} \\
          & \\
         \texttt{omega\_v} & Present-day dark-energy density \cite{Planck:2018vyg} \\
         & \textbf{TYPE:} \texttt{float} \\
          & \\
         \texttt{omega\_m} & Present-day matter density \cite{Planck:2018vyg} \\
         & \textbf{TYPE:} \texttt{float} \\
          & \\
         \texttt{omega\_r} & Present-day radiation density \cite{Planck:2018vyg} \\
         & \textbf{TYPE:} \texttt{float} \\
          & \\
         \texttt{A\_s} & Amplitude of the primordial scalar power-spectrum \cite{Planck:2018vyg} \\
         & \textbf{TYPE:} \texttt{np.float64} \\
          & \\
         \texttt{f\_cmb} & CMB pivot-scale (Hz) \cite{Planck:2018vyg} \\
         & \textbf{TYPE:} \texttt{float} \\
          & \\
         \texttt{gev\_to\_hz} & Conversion from GeV to Hz \\
         & \textbf{TYPE:} \texttt{np.float64} \\
          & \\
         \texttt{g\_rho\_0} & Present-day number of relativistic degrees of freedom \\
         & \textbf{TYPE:} \texttt{np.float64} \\
          & \\
         \texttt{g\_rho\_0} & Present-day number of entropic relativistic degrees of freedom \\
         & \textbf{TYPE:} \texttt{np.float64} \\
     \end{tabular}
     \caption{Constants and conversion factors included in the \texttt{models\_utils} module.}
     \label{tab:const}
\end{center}}
\end{table}
\clearpage
\bibliographystyle{apsrev4-1}
\bibliography{bibliography.bib}
\end{document}